\documentclass{llncs}

\usepackage{llncsdoc}
\usepackage{graphicx}

\newcommand{\keywords}[1]{\par\addvspace\baselineskip
\noindent\keywordname\enspace\ignorespaces#1}

\begin {document}
\title{Improving SAT Solvers via Blocked Clause Decomposition}
\titlerunning{Improving SAT Solvers via BCD}

\author{Jingchao Chen}
\institute{School of Informatics, Donghua University \\
2999 North Renmin Road, Songjiang District, Shanghai 201620, P. R.
China \email{chen-jc@dhu.edu.cn}}

\maketitle
\begin{abstract}

The decision variable selection policy used
by the most competitive CDCL (Conflict-Driven Clause Learning) SAT solvers is either VSIDS (Variable State Independent Decaying Sum) or its variants such as  exponential version EVSIDS. The common characteristic of VSIDS and its variants is to make use of statistical information in the solving process, but ignore structure information of the problem. For this reason, this paper modifies the decision variable selection policy, and presents a SAT solving technique based on BCD (Blocked Clause Decomposition). Its basic idea is that a part of decision variables are selected by VSIDS heuristic, while another part of decision variables are selected by blocked sets that are obtained by BCD. Compared with the existing BCD-based technique, our technique is simple, and need not to reencode CNF formulas. SAT solvers for certified UNSAT track can apply also our BCD-based technique. Our experiments on application benchmarks demonstrate that the new variables selection policy based on BCD can increase the performance of SAT solvers such as abcdSAT. The solver with BCD solved an instance from the SAT Race 2015 that was not solved by any solver so far. This shows that in some cases, the heuristic based on structure information is more efficient than that based on statistical information.


\end{abstract}

\section{Introduction}

The SAT solving technology has made great progress in recent years.
A number of state-of-the-art SAT solvers have come to the fore.
Nevertheless, a great number of SAT problems remain unsolved yet.
How to improve further SAT solvers is still a very important topic.

Recently, Balyo et al \cite{EagerMover:14} reported that reencoding CNF
(Conjunctive Normal Form) formulas by
BCD (Blocked Clause Decomposition) can improve the performance of the state-of-the-art  CDCL (Conflict-Driven Clause Learning) SAT solvers such as Lingeling
\cite{Lingeling:13}. The basic idea behind this technique is to define multiple versions for each variable
through reencoding the original CNF formula. The version number of variables depends on blocked subsets obtained by
a decomposition algorithm. A blocked set is defined to be a set of clauses that can be removed
 completely by BCE (Blocked Clause Elimination)~\cite{BCE:12}. It is easy to verify that
any CNF formula can be decomposed into two blocked subsets. Due to this property, any CNF formula can be reencoded into a new
CNF formula in the order where clauses occur in blocked subsets. According to the experiments of Balyo et al~\cite{EagerMover:14},
some reencoded application benchmarks are indeed easier to be solved than the original ones.  For a part of hard application benchmarks, the reencoding
starts to pay off after 3500 seconds. The main drawback of the reencoding technique is that UNSAT problems solved by it
cannot be certified, since there is no automatic method for recognizing whether the original benchmark and the reencoded benchmark are identical or not. Therefore, the BCD-based reencoding is not suitable for SAT solvers that are used for the certified UNSAT track.

   This paper aims at improving the performance of SAT solvers by BCD. In general, a CDCL SAT solver consists of components such as decision variable selection, Boolean constraint propagation, learnt clause database reduction, restart etc. This paper focuses how to improve the decision variable selection policy of CDCL solvers. Unlike the reencoding approach by Balyo et al~\cite{EagerMover:14}, we do not reencode CNF formulas, but use blocked sets to guide the selection of a few decision variables. Up to now, the decision variable selection policy used by the most competitive CDCL SAT solvers is based on either VSIDS (Variable State Independent Decaying Sum)~\cite{Chaff:5} or its variants such as EVSIDS (exponential VSIDS)~\cite{EVSIDS_A,EVSIDS_B}, VMTF (variable move-to-front), ACIDS (average conflict-index decision score)~\cite{ACIDS}. The common characteristic of VSIDS and its variants is to make use of statistical information in the solving process, but ignore structural information of the problem.
   The empirical evaluation of Biere et al ~\cite{ACIDS} shows that EVSIDS, VMTF, and ACIDS empirically perform equally well. Therefore, we believe that improving the performance of SAT solvers by only modifying VSIDS leads difficultly to a breakthrough.  For this reason,  we decide to use structure information obtained by BCD to
   optimize the decision variable selection policy.  Our basic idea is that a part of the decision variables are selected by VSIDS heuristic, while another part of the decision variables are selected by blocked subsets that are obtained by BCD. Solver abcdSAT~\cite{abcdSAT}, which is built on the top of Glucose 2.3~\cite{LDB,glucose:2.3}, is the first to optimize the variable selection policy with BCD, and  won the Gold Medal of the main track of SAT Race 2015 \cite{sr15}. From the result of SAT
Race 2015, the BCD-based variable selection policy improved indeed the performance of this solver. This paper identifies further such an evaluation. The original abcdSAT applied the BCD-based technique only for small instances, not for large instances. Thus, at SAT
Race 2015, no solver solved a large instance $group\_mulr$. However, if the BCD-based technique is used also for large instances in the initial phase of search,
$group\_mulr$ can be solved in 105.2 seconds by abcdSAT. Our BCD-based technique need not reencoding. Its advantage is that solvers entering certified UNSAT track can apply directly it also.

\clearpage

\section{Preliminaries}

In this section, after the definition of some notations, we introduce the basic principle of a modern CDCL SAT solver, on which the improvement will be made
in Section 4.

A formula in CNF is defined as a conjunction of
clauses, where each clause is a disjunction of literals, each
literal being either a Boolean variable or its negation.
 Usually, the logic form of a clause $C$ is written as $C = x_1 \vee
\cdots \vee x_m$, where $x_i (1 \leq i \leq m)$ is a literal.
A clause with only one literal is called a unit clause or unit literal.
A CNF formula $F$ is written as $F = C_1 \wedge \cdots \wedge C_n $, where
$C_i (1 \leq i \leq n)$ is a clause.

Given two clauses $C_1 = l \vee a_1
\vee \cdots \vee a_m$ and $C_2 = \bar{l} \vee b_1 \vee\cdots \vee
b_n$, the clause $C = a_1 \vee \cdots \vee a_m \vee b_1 \vee \cdots
\vee b_n$ is called the resolvent of $C_1$ and $C_2$ on the literal
$l$,  which is denoted by $C = C_1 {\otimes}_l C_2$.

   The so-called blocked clause can be defined formally as follows.
   Given a CNF formula $F$, a clause $C$,
a literal $l \in C$ is said to block $C$ w.r.t. $F$ if (i) $C$ is a
tautology w.r.t. $l$, or (ii) for each clause $C' \in F$ with
$\bar{l} \in C'$, the resolvent $C' {\otimes}_l C$ is a tautology.
When $l$ blocks $C$ w.r.t. $F$, the literal $l$ and the clause $C$
are called a blocking literal and a blocked clause, respectively.

In general, a CDCL SAT solver consists of unit propagation, variable activity based heuristic, literal polarity phase, clause learning, restarts and a learnt clause database reduction policy etc. Here is the core framework of a CDCL SAT solver.

\begin{small}
\begin{flushleft}
{\bf Algorithm} CDCL\_solver\\
\hskip 4mm   {\bf repeat} the following steps\\
\hskip 12mm   {\bf if} $!propagate()$ {\bf then}\\
\hskip 20mm      {\bf if} $(c = conflictAnalyze())==\emptyset$ {\bf then} return UNSAT\\
\hskip 20mm       add $c$  to learnt clause database\\
\hskip 20mm       backtrack to the assertion level of $c$\\
\hskip 12mm   {\bf else} {\bf if} $(l=pickBranchLit())==null$ {\bf then} return SAT\\
\hskip 20mm              assert literal $l$ in a new decision level\\
\end{flushleft}
\end{small}

In the above algorithm, Procedure $propagate$ performs unit propagation, i.e., assigns repeatedly each unit literal
to true until the formula $F$ has no unit clause under the current model. When this procedure
yields a conflict, a new asserting clause $c$ is derived by Procedure $conflictAnalyze$. If
the derived clause $c$ is empty, then the formula $F$ is unsatisfiable. Otherwise, it is added
to the learnt clause database, and the algorithm backtracks to the assertion level of
the learnt clause $c$, i.e., the level where the learnt clause becomes unit.
If unit propagation does not generate the empty clause, Procedure $pickBranchLit$
begins to select a new decision literal $l$. If such a literal is selected successfully, it is asserted in a new decision level. Otherwise,
the formula is answered to be satisfiable. Throughout this paper, Procedure $pickBranchLit$ is assumed to use the EVSIDS heuristic to select
a literal with the highest score.

\section{Related Work}

In theory, any CNF formula can be decomposed into two blocked subsets
such that both can be solved by BCE (Blocked Clause Elimination).
Therefore, we can assume that one blocked set of a CNF formula
 is $ C_1 \wedge C_2 \wedge \cdots \wedge C_m$, where $C_i$ ($i=1,2,\ldots,m$) is a clause.
Balyo et al \cite{EagerMover:14} reencode each clause $C_i$ so that
the reencoded  formula is solved more easily than the original formula.
Their reencoding may be described as follows. Each blocking literal $x_i$ is allowed to have several versions.
In the order of $C_m, C_{m-1} \ldots , C_1$, each of its versions is defined. Assuming that clause $C_t$ has the form
of $C_t = x_i \vee y_{j_1} \vee \cdots \vee y_{j_k}$, where $x_i$ is
the blocking literal. Let $x_{i,\$}$ be the current version of
$x_i$. Its next version is defined as follows.

\begin{center}
\begin{math}
  x_{i,\$+1}:= x_{i,\$ } \vee ( y_{j_1,\$}  \wedge \cdots \wedge y_{j_k,\$})
\end{math}
\end{center}

\noindent Then this formula is converted to CNF. Clearly, since a blocking literal is mapped to multiple version variables, such a reencoding
technique will add a vast amount of auxiliary variables. As far as we know, so far no solver entering SAT competition (Race) used such a
reencoding technique.

\section{Decision Variable Selection Policy Based on Blocked Clause Decomposition}

In this section we describe a new SAT solving technique, which is based on BCD. 

  Apart from the reencoding technique of Balyo et al, to avoid using auxiliary variables, our BCD-based solving
 policy does not reencode the original CNF formula, but uses blocked subsets to guide the selection of decision variables.
 The reencoding technique of Balyo et al is to how to convert a CNF formula into one which is easily solved, but does not modify
 any SAT solver. Nevertheless, our BCD-based solving technique is to modify
 a SAT solver.
 Assuming that a CNF formula is decomposed into a large blocked subset $L$ and a small blocked subset $S$. The reencoding
 technique uses each blocked subset separately, while we use them by appending $S$ to $L$. For convenience, let $L \wedge S =
  C_1 \wedge C_2 \wedge \cdots \wedge C_n$, where $C_i$ ($i=1,2,\ldots,n$) is a clause, and the order of clauses in $L$ and $S$ is opposite to
that in which clauses are eliminated by BCE. We use the locations where variables occur in $L \wedge S$ for the first time to determine
decision variable selection at some levels in the pickBranchLit procedure of a CDCL solver. Let $pos[v]$ denote the minimum index where
variable $v$ occur in $C_1 \wedge C_2 \wedge \cdots \wedge C_n$.
Ignoring the priority of clauses, the position indexes of variables may be computed as follows.

\begin{small}
\begin{flushleft}
\hskip 4mm   {\bf for } each variable $v$ {\bf do} $pos[v]=0$ \\
\hskip 4mm   {\bf for } $i = 1$ to $n$ {\bf do}\\
\hskip 12mm     {\bf for} each variable $v \in C_i$ {\bf do}\\
\hskip 20mm         {\bf if} $pos[v]=0$ {\bf then} $pos[v]=i$\\
\end{flushleft}
\end{small}

\noindent In our real implementation, the priority of binary clauses is higher than
that of the other clauses. That is to say, if there exist non-binary clause $C_j$ and binary clause $C_k$ that
both contain $v$, and there is no binary clause $C_i$ ($i<k$) containing $v$,  $pos[v]$ is set to $k$ even if $j < k$.
Hence, the exact expression of $pos[v]$ may be described as follows.
\[{\it pos(v)}= \left \{ \begin {array}
         {l@{\quad \quad}l}
         \arg \min \{v \in C_i \wedge |C_i|=2\} &  \exists_i |C_i|=2 \\
         \arg \min \{v \in C_i\} &  \forall_i |C_i|\neq 2 \\
         0 & \mathrm {otherwise}
         \end {array} \right . \]\\
 Using $pos[v]$, procedure \emph{pickBranchLit} of CDCL SAT solvers may be modified as follows.

\begin{small}
\begin{flushleft}
{\bf Algorithm} pickBranchLit\\
\hskip 4mm   {\bf if} current level $ \in \{1,2,3\} \wedge \#conf < \theta $ {\bf then}\\
\hskip 12mm     Let $v$ be decision variable at 0 level \\
\hskip 12mm     $S \leftarrow \emptyset$ \\
\hskip 12mm     {\bf for} $i=pos[v]$ to $pos[v]+5$ {\bf do}\\
\hskip 20mm         {\bf if} $C_i$ is satisfied {\bf then continue}\\
\hskip 20mm          $S \leftarrow S \cup C_i$ \\
\hskip 12mm     {\bf if} $S \neq \emptyset$ {\bf then return} literal $l \in S$ with EVSID-based highest score \\
\hskip 4mm   {\bf return} literal $l \in F$ with EVSID-based highest score
\end{flushleft}
\end{small}

\noindent At whichever decision level, the algorithm selects always a literal $l$ with the highest score computed by EVSID.
At the 1st -- 3rd decision level, the selection range is limited to $S$, which is a subset of formula $F$,
while at the other levels, it is $F$, not subset $S$, i.e., there is no limitation.
According to our experiments, it is a good choice that only those three levels adopt the BCD-based decision variable selection policy.
In the above pseudo-code, $\#conf$ denote the number of conflicts. In general, $\theta $ is set to 30000 for large instances, and 500000 for small
instances. We use condition $\#conf < \theta $ to limit the application range of the BCD-based policy.
When selecting a decision variable at the 1st -- 3rd level,
we consider at most 6 clauses $C_i$ ($pos[v] \leq i \leq pos[v]+5$) in the order of the blocked clauses,
where $v$ is a decision variable at the 0-th level.
Among these candidate clauses, we pick a literal with the EVSID-based highest score as
a decision literal. Whenever a part of variables are fixed, our solver runs a simplification procedure. In general, after the simplification procedure, we can obtain a blocked set that is different from the initial blocked set. However, to save the cost of computing repeatedly the blocked set, we do not update $pos[v]$.
In other words, what we used  is the initial $pos[v]$ (blocked set), not the updated $pos[v]$.
From this viewpoint, our BCD-based policy is static, not dynamic.

\section{Empirical evaluation}

   We evaluated the performance of SAT solving with BCD and without BCD under the following
experimental platform: Intel Core 2 Quad Q6600 CPU with speed of
2.40GHz and 2GB memory.



\begin{table*}
\caption{Runtime of solver abcdSAT with different modes on application benchmarks that were not solved by at least one out of four modes: no BCD, BCD1-3.
 $|F|$ is in thousands of clauses. Time is in seconds.}
\begin{center}

\setlength\tabcolsep{4pt}
\begin{tabular}{|l|c|c|c|c|c|c|c|}

\hline  \hline
 \multicolumn{1}{|c|}{Instances} & $|F|$ & $\#var$ & S/ & \scriptsize{abcdSAT} & \scriptsize{abcdSAT} & \scriptsize{abcdSAT} & \scriptsize{abcdSAT}\\
                                 &       &         & U  & \scriptsize{no BCD}  & \scriptsize{ BCD1}   & \scriptsize{ BCD2}   & \scriptsize{ BCD3} \\
 \hline
korf-18                           & 207  & 7178    & U  & $>5000$ & 420.5   & 579.8   & 420.5 \\
group\_mulr                       & 4302 & 1052071 & U  & $>5000$ & $>5000$ & 105.2   & 105.2 \\
52bits\_12.dimacs                 & 19   & 872     & S  & $>5000$ & $>5000$ & 4940.5  & 4940.5 \\
aes\_32\_3\_keyfind\_1            & 2    & 397     & S  & 180.1   & $>5000$ & 4324.8  & 4324.8  \\
aes\_64\_1\_keyfind\_1            & 2    & 276     & S  & $>5000$ & 846.7   & 2457.7  & 2457.7 \\
grieu-vmpc-31                     & 104  & 961     & S  & 1566.6  & 4966.0  & $>5000$ & 1566.6 \\
gss-22-s100                       & 52   & 9330    & S  & $>5000$ & 812.6   & $>5000$ &  812.6 \\
\scriptsize{jgiraldezlevy.2200.9086.08.40.2}   & 11  & 1998 & S  & $>5000$ & 608.9   & 4148.6  & 608.9 \\
\scriptsize{jgiraldezlevy.2200.9086.08.40.62}  & 11  & 2048 & S  & $>5000$ & 829.7   & 1303.2  & 829.7 \\
\scriptsize{jgiraldezlevy.2200.9086.08.40.83}  & 11  & 1988 & S  & 2038.1  & $>5000$ & $>5000$ & $>5000$ \\
\scriptsize{jgiraldezlevy.2200.9086.08.40.93}  & 12  & 2039 & S  & $>5000$ & 4055.7  & 864.2   & 4055.7 \\
\scriptsize{manthey\_DimacsSorter\_35\_8}      & 52  & 3349 & S  & $>5000$ & 2841.8  & 3591.8  & 2841.8 \\
\scriptsize{manthey\_DimacsSorterHalf\_35\_8}  & 52  & 3349 & S  & $>5000$ & 2729.1  & 3448.7  & 2729.1 \\
\scriptsize{manthey\_DimacsSorter\_37\_3}      & 57  & 3492 & S  & 426.1   & $>5000$ & $>5000$ & $>5000$ \\
\scriptsize{manthey\_DimacsSorterHalf\_37\_3}  & 57  & 3492 & S  & 439.2   & $>5000$ & $>5000$ & $>5000$ \\
mrpp\_8x8\#22\_24                              & 118 & 10097& S  & $>5000$ & 4378.1  & $>5000$ &4378.1  \\
\scriptsize{manthey\_DimacsSorterHalf\_37\_9}  & 70  & 4594 & S  & 2824.5  & $>5000$ & $>5000$ & $>5000$ \\
\hline

\end{tabular}
\end{center}
\end{table*}

\begin{figure}
\centering
\includegraphics[height=7.2cm]{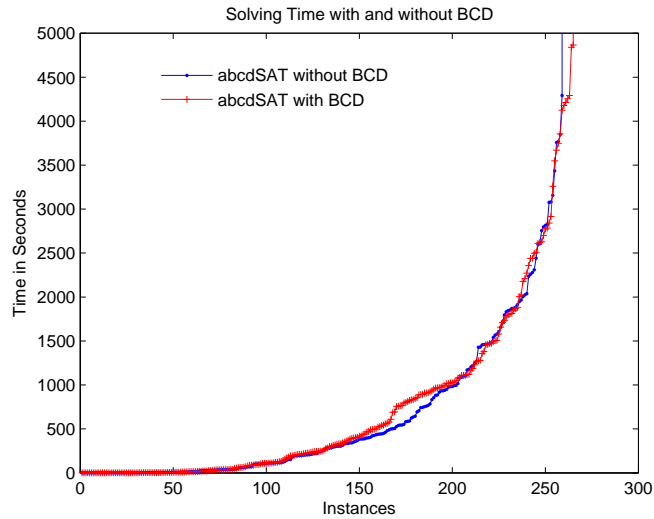}
\caption{The number of instances that abcdSAT with and without BCD can
solve in a given amount of time. The x- and y-axis denote the
number of solved instances and running time in
seconds, respectively. The time limit for each instance was 5000 seconds.} \label{Fig1}
\end{figure}

In this experiment, to verify the efficiency of the BCD-based policy, we use solver abcdSAT \cite{abcdSAT},
the winner of the main track of SAT Race 2015, which is built on the top of the CDCL solver Glucose 2.3. BCD algorithm used
is a recently proposed algorithm called \emph{MixDecompose}~\cite{MixBCD},
The source code of \emph{MixDecompose} is available at
http://github.com/jingchaochen/MixBcd.
As an advantage, \emph{MixDecompose} can ensure that the decomposition of any instance is done within 200 seconds.
And its decomposition quality keeps high still. Here, the decomposition quality is measured by $\frac{|L|}{|F|}$, where $|L|$ and $|F|$
denote the number of clauses in the large blocked set and the original formula, respectively.

Table\,1  shows the runtime of abcdSAT with four different modes (versions).
The timeout for each solver to solve each instance was set to 5000 seconds. Mode \emph{no BCD}
denotes that abcdSAT doest not use any BCD-based policy. All the other three modes are  based on BCD.
The difference among these three modes is that the values of $\theta $ in condition $\#conf < \theta $ in algorithm \emph{pickBranchLit}
are different. Let $n$ and $m$ denotes the number of clauses and variables in formula $F$,
respectively. $\theta $ is set as follows.\\
\vskip -1.5mm
\noindent Mode {\it BCD}1: \[\theta= \left \{ \begin {array}
         {l@{\quad \quad}l}
         0 & n > 15 \times 10^5 \vee m > 5\times 10^5\\
         6 \times 10^6 & \mathrm {otherwise}
         \end {array} \right . \]\\
Mode {\it BCD}2: \[\theta= \left \{ \begin {array}
         {l@{\quad \quad}l}
         0 & n > 5\times10^6 \vee m > 15 \times 10^5 \vee n < 2m\\
         30000 & \textrm {above is false} \wedge m > 5 \times 10^5 \\
         5\times 10^5 & \mathrm {otherwise}
         \end {array} \right . \]\\
Mode {\it BCD}3: \[\theta= \left \{ \begin {array}
         {l@{\quad \quad}l}
         0 & n\mathrm{>}5\mathrm{\times}10^6\mathrm{\vee}m\mathrm{>}15\mathrm{\times}10^5\mathrm{\vee}n\mathrm{<}2m\mathrm{\vee}n\mathrm{>}30m\\
         30000        &  \textrm {above is false} \wedge m > 5 \times 10^5 \\
         6\mathrm{\times}10^6 &  \textrm {above is false} \wedge m \geq 1600 \wedge m \leq 15000 \\
         5\mathrm{\times}10^5 &  \mathrm {otherwise}
         \end {array} \right . \]\\
Mode \emph{BCD}1 is actually the SAT-Race 2015's version of abcdSAT.

In Table\,1, Column $|F|$ denotes the number of clauses
in formula $F$ in thousands of clauses.  Here $F$ is
a formula simplified by the preprocessing of abcdSAT, not the original input formula.
$\#var$ denotes the number of variables in $F$.
Column $S/U$ indicates whether an instance is SAT or UNSAT.
Table\,1 lists all the application benchmarks from SAT Race 2015 where the performance of four modes is inconsistent,
except the first one from SAT Competition 2014. For benchmarks that are not listed in Table\,1,  either all the four modes solved them or no
mode solved them in 5000 seconds. As seen in Table\,1,  the solvers with BCD are better than the solver without BCD. The performance of Mode
\emph{BCD}3 is the best. It solved 7 more instances than the mode without BCD. Mode \emph{BCD}3 solved $group\_mulr$ in
105.2 seconds. This instance was not solved by any solver in the SAT Race 2015. Notice, the SAT-Race 2015's version of abcdSAT adopted the BCD policy
only for small instances, not for large instances such as $group\_mulr$. In addition, no MiniSat-style solver solved \emph{korf-18}. Our BCD-based version solved easily it.

Figure \ref{Fig1} shows a cactus plot related to the comparison of abcdSAT with and without BCD.
In this comparison experiment, all 300 instances tested are from the main track at the SAT Race 2015.
Here the BCD mode used is Mode \emph{BCD}3. As seen in the cactus plot,
when the given amount of time is small, the solver with BCD has no advantage. However, when it
is plenty large enough, the solver with BCD solved more instances than the solver without BCD.

\section {Conclusions}

  In this paper, we  presented a BCD-based improvement strategy on SAT solvers, which is different from
  that of Balyo et al \cite{EagerMover:14}. Common to the two strategies is that they start to pay off
  after sufficiently long time. If a given amount of time is very short, there is no improvement on SAT solvers.
  Compared with the approach of Balyo et al, our approach is simple, need not reencoding, and has no application limit.
  In addition, the BCD-based reencoding of Balyo et al is a preprocessing, while our BCD-based mode is a solving strategy.

  The decision variable selection is a very important component of CDCL SAT solvers. Our BCD-based variable selection policy solved an instance that was not solved by any solvers so far. However, it seems to be suited only for a part of instances. What is the optimal variable selection policy after all? This is a problem which is well worth looking into further. The current version of our BCD-based strategy is static. However, in the usual sense, dynamic is better than static. It is left as an open problem how to make the BCD-based strategy dynamic.

\bibliographystyle{splncs}
\bibliography{satBCD}

\begin{thebibliography}{[MT1]}
%
\bibitem{LDB}
Audemard, G.,  Simon, L.: Predicting learnt clauses quality in modern sat solvers,
Proceedings of IJCAI 2009, pp. 399--404 (2009)

\bibitem{glucose:2.3}
Audemard, G.,  Simon, L.: Glucose 2.3 in the sat 2013 competition,
Proceedings of the SAT Competition 2013, pp.40--41 (2013)

\bibitem{EagerMover:14}
Balyo, T.,  Fr\"{o}hlich, A., Heule, M.J.H., Biere, A.: Everything you always wanted to know
about blocked sets (but were afraid to ask), SAT 2014, LNCS 8561, pp. 317--332 (2014)

\bibitem{ACIDS}
Biere, A., Fr{\" o}hlich, A.: Evaluating cdcl variable scoring schemes,
SAT 2015, LNCS 9340, pp.405--422 (2015)

\bibitem{EVSIDS_B}
Biere, A.: Adaptive restart strategies for conflict driven sat solvers,
SAT 2008, LNCS 4996, pp. 28--33 (2008)

\bibitem{Lingeling:13}
Biere, A.: Lingeling, plingeling and treengeling entering the sat competition 2013,
Proceedings of SAT Competition 2013, pp. 51¨C-52. University of Helsinki (2013)

\bibitem{MixBCD}
Chen, J.C.: Fast blocked clause decomposition with high quality, 2015, http://arxiv.org/abs/1507.00459.

\bibitem{abcdSAT}
Chen, J.C.:
MiniSAT\_BCD and abcdSAT: solvers based on blocked blause decomposition,
Proceedings of SAT Race 2015.

\bibitem{EVSIDS_A}
E{\' e}n, N.,  ~S{\" o}rensson, N.: An extensible sat-solver,
SAT 2003, LNCS 2919, pp. 502--518 (2004)

\bibitem{BCE:12}
J\"{a}rvisalo, M., Biere, A., Heule, M.J.H.: Simulating circuit-level simplifications on
cnf, Journal of Automated Reasoning 49(4), pp.583--619 (2012)

\bibitem{Chaff:5}
Moskewicz, M.W., Madigan, C.F., Zhao, Y., Zhang, L.T., Malik, S.:
Chaff: Engineering an Efficient SAT Solver. Design Automation
Conference (DAC) (2001)

\bibitem{sr15} SAT-Race 2015 web page: http://baldur.iti.kit.edu/sat-race-2015/

\end{thebibliography}

\end{document}